\documentclass{article}

\usepackage[english]{babel}

\begin{document}

\title{Solitons as Key Parts to Produce a Universe in the Laboratory}

\date{14th September 2006}

\author{Stefano Ansoldi\footnotemark[1]$\quad$and Eduardo I. Guendelman\footnotemark[2]}
\footnotetext[1]{\textit{Universit\`{a} degli Studi di Udine, Udine, Italy}, \textrm{email:} \texttt{ansoldi@trieste.infn.it}}
\footnotetext[2]{\textit{Ben Gurion Univeristy, Beer Sheva, Israel},  \textrm{email:} \texttt{guendel@bgu.ac.il}}

\maketitle

\begin{abstract}
Cosmology is usually understood as an
observational science, where experimentation
plays no role.
It is interesting, nevertheless, to change this
perspective addressing the following question:
what should we do to create a universe, in a
laboratory? It appears, in fact, that this is,
in principle, possible according to at least
two different paradigms; both allow
to circumvent singularity theorems,
i.e. the necessity of singularities in the past
of inflating domains which have the required properties
to generate a universe similar to ours.

The first of them is substantially classical, and
is built up considering solitons which collide
with surrounding topological defects, generating
an inflationary domain of space-time.

The second is, instead, partly quantum and
considers the possibility of tunnelling of
past-non-singular regions of spacetime into an
inflating universe, following a well-known
instanton proposal.

We are, here, going to review some of these models,
as well as highlight possible extensions, generalizations
and the open issues (as for instance the detectability of
child universes and the properties of quantum tunnelling
processes) that still affect the
description of their dynamics. In doing so we will
remark how the works on this subject can represent
virtual
laboratories to test the role that fundamental principles of
physics (particularly, the interplay of quantum and general
relativistic realms) played in the formation of our universe.

\end{abstract}

\section{State of the Art of Universe Creation in the Laboratory}

Cosmology is usually considered as an observational
discipline, not susceptible
of a direct experimental approach. This seems,
of course, quite intrinsic to
the discipline itself, which deals with problems
related to the birth
of our universe and its evolution in the present
state; in fact the very
meaning of the word \textit{cosmology}
(stemming from the Greek word \textit{cosmos},
which meant \textit{beauty}, \textit{harmony})
seems to bind us to a passive,
contemplative attitude in the study of the universe.

On the other hand, even without considering
phenomena proper of the quantum
world, already General Relativity, a
classical theory, brings
a challenge to the above perspective, by raising
for the first time the
concept of causality as a central one in physics.
Because of causal relationships,
the domain of what can be experienced/observed might be,
or more likely \textit{is},
only a subset of what is existing. At the same
time the very
concept of causality and its relationship with the
global spacetime structure
brings up another problem in the scenario of cosmology.
In fact, the simplest models
of the universe built according to General Relativity
and with an at least
reasonable degree of consistency with what we observe,
seem doomed to begin with
an initial singularity. The breakdown of field equations
at the very first event
of our history (universe creation) is very unfortunate
for our description in
terms of differential equations, since field equations
break down exactly where
we would like to set up the initial conditions.

At the same time, if we trace the life of
our universe back in time closer and
closer to its very beginning, for instance
at the exit of the inflationary era,
we will see that many parameters
describing it are quite far from the
domain of ``very large scales'' which characterizes
the present observable universe. In fact
for a Grand Unified Theory scale of
$10 ^{14}$ Gev the universe could emerge
from a classical bubble which starts
from a very small size if the mass of
the bubble is of the order of
about $10$ Kg (by using quantum tunnelling
the mass of the bubble could be
arbitrarily small, but the probability of
production of a new universe out
of it would be reduced).
Although the density of the universe would
have been quite higher than
what we can realize with present technologies,
the orders of magnitude of the
other parameters make not unreasonable to
ask the question: might we be
able to build a universe in the laboratory?
Already years ago it was, in fact,
recognized that this question can have,
in principle, a positive answer and a simple
model of the creation process involving
semiclassical effects was suggested.
Since then, further decisive developments
of those early results have not
appeared; there have been a few more proposals,
addressing with simple models some
qualitative issues, but the problems which emerged
in the earliest formulations are,
somehow, still open. In our opinion, it is
certainly interesting, if not necessary,
to study in more detail and with systematic
rigor these problems as well as other
realistic answers to the above question.
As we pointed out, this question is not a purely
academic one and we think that a different
perspective (an experimental rather than
observational one) has interesting consequences.
In the rest of this section we are then going
to give a concise review of the state of the
art in the field according to this point of view.
It is worthwhile to remark that, this perspective
can be considered much more promising nowadays,
since observations have pushed our eyes further
back in time, providing us
with information about our universe in its
earliest stages of life. This has
allowed tighter constrains on the parameters of
models describing the early universe,
so that most of the problems can now be attacked
more easily.
For example, it is now easier to identify the
fundamental elements (building blocks)
which we can use to model the creation of a
universe that will evolve in something similar to
the present one. At the same time, we could have
a deeper insight of the fundamental
principles that forged the earliest evolution of
the universe, with more hope to enlighten
a crucial one, which is the interplay between
General Relativity and Quantum Theory.

This said, we are now going to make a closer contact
with some of the models for the formation
of the universe.
These are the ones studying the dynamics of
vacuum bubbles and those developing
the idea of topological inflation
(both aspects considered also
in a semiclassical framework).

The study of vacuum decay initiated more than
30 years ago with the work of Callan and Coleman
\cite{bib:PhReD1977..15..2929C,bib:PhReD1977..16..1762C};
in the following
years the interest in the subject increased and the
possible interplay of true vacuum
bubbles with gravitation was also studied
\cite{bib:PhReD1980..21..3305L,bib:PhReD1987..36..1088W}.
At the same time, and as opposed with the true vacuum
bubbles of Coleman \textit{et al.}, false
vacuum bubbles were also considered. In connection
with gravity, the classical
behavior of regions of false vacuum,
first studied by Sato et \textit{al.}
\cite{bib:PrThP1981..65..1443M,%
bib:PrThP1981..66..2052M,bib:PrThP1981..66..2287S,%
bib:PhLeB1982.108....98K,bib:PhLeB1982.108...103M,%
bib:PrThP1982..68..1979S}, was
for example analyzed in
\cite{bib:PhReD1987..36..2919T,bib:PhReD1987..35..1747G}
and in \cite{bib:PhReD1991..43.R3112T}.

From the classical point of view, false vacuum
bubbles have an energy density which is higher
than that of the surrounding spacetime.
Because of this, although the space inside
the bubble can undergo an exponential inflation,
the pressure difference with respect to
the outside implies that the bubble cannot
displace the external space.
The \emph{child universe} solutions appear
then as expanding bubbles of false vacuum
which disconnect from the exterior region.
These solutions contain a wormhole and
also a \emph{white-hole like} initial singularity.
They are allowed under general relativistic settings,
where, in the simplest case,
the region inside the bubble can be modelled by a
domain of de Sitter spacetime and the
outside by a domain of Schwarzschild spacetime;
these two regions are then joined across the bubble
surface, using the well known junction conditions
\cite{bib:NuCim1966.B44.....1I,bib:NuCim1967.B48...463I,%
bib:PhReD1991..43..1129I};
on the bubble Einstein equations, which also hold
separately in the two domains, are
satisfied when interpreted in a distributional
way and determine the motion (embedding)
of the bubble itself in the two domains of spacetime.
Although there are configurations of this system
(and of more elaborate generalizations)
that are appropriate to describe the evolution
of a newly formed universe (in the sense
that the expanding bubble can become very large)
these models are affected by some pathologies.
In particular:
\begin{enumerate}
    \item[$-$] only bubbles with masses above
    some critical value can expand from very small
    size to infinity;
    \item[$-$] all the solutions, which can expand
    enough to represent a new universe starting
    from a very small size, have a singularity (white-hole) in their past, since, for them, all
    hypotheses of singularity theorems are satisfied.
\end{enumerate}

In connection with the first problem it was observed in
\cite{bib:PhReD1991..44..3152R}\footnote{{The
subject of inflation assisted by topological
defects was also studied later in
\cite{bib:PhReL1994..72..3137V} and
\cite{bib:PhLeB1994..327...208L}.}}
that in theories with multiple scalars,
like a triplet of scalars, all bubbles that
start evolving from zero radius can inflate
to infinity, provided the scalars are in
a ``hedgehog'' configuration or global monopole
of big enough strength. This effect also
holds in the gauged case for magnetic monopoles
with large enough magnetic charge.

A possible connection of this with the second
problem mentioned above, appears from
the discussion of Borde \textit{et al.}
\cite{bib:PhReD1999..59043513V}: they
proposed the possibility of a mechanism
by means of which two regular magnetic
monopoles (with \textit{below critical}
magnetic charge) could coalesce and form
a \textit{supercritical} one, which then
inflates and gives rise to a child universe.
This process might help addressing the
singularity problem. In this context it is very
interesting the work of Sakai \textit{et al.}
\cite{bib:gr-qc2006..02...084K} in which
the interaction of a magnetic monopole
with a collapsing surrounding membrane
is considered; also in this case a
new universe can be created.

Related to the solution of this second issue
are a number of other approaches that propose to
use quantum effects. In particular when
describing the bubble separating the inflating
spacetime domain from the surrounding spacetime
in terms of Israel junction conditions
\cite{bib:NuCim1966.B44.....1I,bib:NuCim1967.B48...463I,%
bib:PhReD1991..43..1129I} (and under the additional assumption
of spherical symmetry) it is possible to reduce the
problem to the study of a system with only one degree
of freedom: this is the so called \textit{minisuperspace
approximation}, which has been adopted to address the problem
of the semiclassical quantization of the system even
in the absence of an underlying
quantum gravity theory. This has been the approach
by\footnote{Apart from the papers already cited above,
the semiclassical approach has also been discussed
by other authors (see for instance
\cite{bib:PhReD1997..56..4651E,bib:PhReD1997..56..4663E,%
bib:ClQuG1997..14...L59M}) and we would also like to recall
the suggestive relationship between
the decay of the cosmological
constant, membranes generated by higher rank gauge
potentials and black holes, which
have appeared in many papers in the literature
\cite{bib:NuPhy1980B176...509T,bib:PhLeB1984.143...415T,%
bib:PhLeB1987.195...177T,bib:NuPhy1988B297...787T,%
bib:ClQuG1989...6..1379M,bib:NuPhy2001B602...307W,%
bib:PhReD2001..64025008S,bib:PhReD2004..69083520W,%
bib:IJThP2004..43...883M}.}
Farhi \textit{et al.}
\cite{bib:NuPhy1990B339...417G} and by Fishler \textit{et al.}
\cite{bib:PhReD1990..41..2638P,bib:PhReD1990..42..4042P}.
One difficulty that these approaches faced was that the initial
state was not a classically stable one. This was resolved
by the introduction of massless scalars or gauge fields
that live on the shell and produce classical stabilization
of false vacuum bubbles.
These bubbles can then, by quantum tunnelling, become child universes
\cite{bib:ClQuG1999..16..3315P}. In a $2+1$-dimensional example
\cite{bib:MPhLA2001..16..1079P}
the tunnelling can be arbitrarily small.

\section{Outlook and Prospects for Future Research}

Most of the models that have been developed
to describe the process of
universe creation are based on a very well-known
and studied classical system,
usually known as a \textit{general relativistic shell}
\cite{bib:NuCim1966.B44.....1I,bib:NuCim1967.B48...463I,%
bib:PhReD1991..43..1129I}. General relativistic
shells provide an excellent, non-trivial,
gravitational system, whose dynamics
can be described by an extremely intuitive set
of equations with a clear geometrical
meaning. In situations with high symmetry the
number of equations of motion of the
system is drastically reduced (and often we are
left with just one non-linear
equation). In this sense, the classical dynamics
of the system is ``under control'';
there are then many analytical results that can
be derived and numerical methods
have also been employed (see the introduction of
\cite{bib:ClQuG2002..19..6321A} for
additional references). What is, somehow, surprising
is that little progress
has been made in the development of the quantized
theory, which still remains a
non-systematized research field. A progress in
this direction would be very important
to be able to analyze, for instance, the
semiclassical process of universe creation.

Let us first concentrate our attention on the
classical creation process. We especially
remember, in this context, the works
of Borde \textit{et al.} \cite{bib:PhReD1999..59043513V}
and of Sakai \textit{et al.} \cite{bib:gr-qc2006..02...084K},
which suggest many interesting ideas for further
developments in the field.
\begin{itemize}
    \item[$-$] For instance, it would be possible to extend
    the analysis in \cite{bib:PhReD1999..59043513V} which is
    mainly qualitative in nature: in fact, the process of
    collision of two magnetic monopoles and the
    formation, by means of it, of a supercritical one,
    is highly non-linear; the
    detailed analysis of this non-linearity is instrumental
    for a quantitatively
    meaningful use of the idea of topological inflation.
    \item[$-$] Also one could extend the study performed
    in \cite{bib:gr-qc2006..02...084K},
    making a complete analysis of all the possible
    choices of the parameters of the model and
    of the related spacetime structures; this should help
    to draw a definitive conclusion about the classical
    stability of the initial configuration,
    i.e. to determine if this is a general feature of
    monopole models or if it appears only
    for a restricted subset in the full parameter space.
\end{itemize}
In both the above mentioned classical models another
crucial problem is the one
about the causal structure of the resulting
spacetime describing the universe
creation. In fact the global spacetime
structure can be obtained by well-known
techniques. Again a full classification
of all the possibilities that can arise
is certainly necessary to gain evidence in
favor or against the proposed mechanisms.
It is already known that subtleties can appear
in some of these cases, as,
for example, the presence of singularities
in the causal past of the created universe
\emph{but} not in the past of the experimenter
creating the universe in the laboratory.
Also the presence of timelike singularities,
that in some cases are not hidden behind
horizons (i.e. are \emph{naked}), makes
interesting a discussion, in this context, of the
problem of initial conditions. The proper
analysis of the Cauchy problem will, in fact,
involve resolution or proper handling of these
singularities.

After discussing the classical aspects, we now
come to the quantum
(more precisely semiclassical) ones.
Let us first recall that the semiclassical
picture invokes quantum effects to justify
the tunnelling between a classical solution,
that can be formed without an initial
singularity, and another classical solution,
which can describe an inflating universe.
In this way, the creation of the inflating
universe \emph{via} quantum tunnelling,
could evade the consequences of singularity
theorems, i.e. the
initial singularity. A first problem which
has already been partly studied is the
stability of the initial classical configuration
\cite{bib:ClQuG1999..16..3315P}.
It is interesting then to
consider the tunnelling process in more
general situations, where, for example,
the stabilization can be still classical
in origin.
In \cite{bib:gr-qc2006..02...084K}
it is shown that it is possible to obtain
this solution in the context of monopole
configurations, although, as we mentioned
above, the analysis should be extended
to the whole of the parameter space.
At the same time it can also be interesting to
consider the possibility that semiclassical
effects might stabilize the initial
configuration. In particular,
closely related to the problem of
instabilities present in many models,
is the fact that the spacetime surrounding the vacuum
bubble (which is often chosen to be Schwarzschild
spacetime or generalizations of it)
has itself an instability due to presence of the
white hole region (see, for instance,
\cite{bib:PhReD1977..16..3359T}). Also in this
context quantum effects might stabilize the system
and help solving the issue.
A possible suggestion in this direction,
requires the determination of the
stationary states in the
WKB approximation, so we propose to
generalize the procedure presented in
\cite{bib:ClQuG2002..19..6321A} (where
this analysis was performed for the first
time in a simplified model) to the
configurations that we considered above.

Another important point for further
investigations, could be to address with a critical
spirit the issues related with the
semiclassical tunnelling procedure. About this,
we are now going to follow, for definiteness,
the clear, but non-conclusive,
analysis developed by Farhi
\textit{et al.} \cite{bib:NuPhy1990B339...417G}):
they show that, when considering
the tunnelling process, it is not possible
to devise a clear procedure to build the manifold
interpolating between the two (initial and final)
classical configurations; this
manifold would describe the instanton that is
assumed to mediate the process.
In particular, Farhi \textit{et al.} show that
it seems possible to build
only what they call a \emph{pseudo-manifold},
i.e. a manifold in which various points
have multiple covering. To make sense of
this, they are forced to introduce
a `covering space' different from the
standard spacetime manifold, in which
they allow for a change of sign of the
volume of integration required for
the calculation of the tunnelling action
and thus of tunnelling probabilities.
It is suggestive to consider other approaches
which might give a more precise
meaning to the concept of a
\emph{pseudo-manifold}. In this context
we would like to recall two possibilities.
\begin{enumerate}
    \item[$-$] A first one uses the \textit{two measures
    theory} \cite{bib:PhReD1999..60065004K}, where
    one can use four scalar fields and define
    an integration measure in the action
    from the determinant of the mapping between
    these scalar fields and the four
    spacetime coordinates; there can be
    configurations where this mapping is not of
    maximal rank, and this appears to be relevant
    to the problem we are discussing,
    if we interpret the scalar fields as coordinates
    in the \emph{pseudo-manifold}
    of \cite{bib:NuPhy1990B339...417G}. In this
    picture the non-Riemannian volume
    element of the two measure theory would be
    related to the non-Riemannian
    structure that must be associated to the
    \emph{pseudo-manifold}, as recognized by
    Farhi \textit{et al.}. Thus, it appears that
    the consideration of non-Riemannian
    volume elements could be essential to make
    sense of the quantum creation of a universe in
    the laboratory, so that it could be important to develop
    the theory of shell dynamics in the framework described
    by the two measures theory.
    \item[$-$] A second one, likely complementary,
    can come from a closer study of the Hamiltonian
    version of the dynamics of the system.
    To better understand this point, we remember
    that the Hamiltonian for a general relativistic
    shell, which we are using as a model for the
    universe creation process, is a non quadratic
    function of the momentum, because of the
    non-linearities intrinsic to General Relativity;
    this makes non-standard and subtle the
    quantization procedure. Although it is possible
    to determine an expression for the Euclidean
    momentum which reproduces standard
    results for the decay of vacuum bubbles,
    this momentum can have unusual properties
    along the tunnelling trajectory; it turns
    out that some of these inconsistencies
    disappear if we consider the momentum as
    a function valued on the circle instead than on
    the real line \cite{bib:gr-qc2006..xx...zzzS};
    further investigations in this direction
    are then likely to give a better understanding
    of the semiclassical tunnelling creation
    of a universe in the laboratory.
\end{enumerate}

With the above we have recalled some existing
ideas to analyze the creation of a
universe by classical or quantum processes;
we also presented some significative
further developments of these ideas.
Another related question is if
all \emph{creation efforts} might end
in a baby universe totally disconnected from its
creator or not. Since there is not a definitive
answer to this problem yet,
it is certainly interesting to address
the question if, in some way, the new
universe might be detectable. There is an
indication in this direction from the
analysis performed in
\cite{bib:PhReD1991..44...333M}, where a
junction with a Vaidya radiating metric is employed
so that the baby universe would be detectable because
of the modifications to the Hawking radiation due
to the baby universe creation process.
It is natural to think generalizations that apply
to solitonic inspired universe creation,
for instance extending the metric describing
the monopole, i.e.
the Rei\ss{}ner-Nordstr\"{o}m spacetime, to the
Rei\ss{}ner-Nordstr\"{o}m-Vaidya case.
This question can be important in the perspective
of a gravitational scenario where quantum
effects are also relevant and in which the
exact and definite character of the classical
causal relations proper of general relativity, might
be altered by quantum effects.

Once one has some detailed model of universe
creation, many more phenomenological issues
can be analyzed and the differences between
purely classical and semiclassical processes can
also be better appreciated through this analysis.
This is a good reason to consider both
procedures explicitly and separately,
together with the
physical consequences of different values of
the parameters characterizing the newly
forming/formed universe. In particular it
is possible to study how different ways of
creating a universe
in the laboratory could lead to different
resulting universes, with, maybe, different
coupling constants, gauge groups, etc..
In this context we would also like to recall the
hypothesis of Zee \textit{et. al}
\cite{bib:phys.2005..10...102Z}, that a creator
of a universe could pass a message to the future
inhabitants of the created universe. In the perspective
that we gave in this short review,
this is a suggestive way to represent the problem
of initial conditions and of the
causal structure, which can be of relevance
also for the problem of defining probabilities
in the context of the multiverse theory and
of eternal inflation.
Another point of phenomenological relevance,
is the connection between universe creation
and the current observations that suggest
the universe as super-accelerating. It may seem
that, if this result will be confirmed,
it will support the idea that some very unusual
physics could be governing the universe,
in the sense that standard energy conditions
might not be satisfied. The process of
creation of baby universes in the laboratory
without initial singularity deserves closer
investigation, since it might seem plausible
that the basic behavior of the universe to
try to raise its vacuum energy could take a
local form, i.e. manifest itself in the
creation of bubbles of false vacuum (as seen
by the
surrounding spacetime), which would then led
to child universes. A proposal, based on the two
measures theory, for avoiding initial
singularities in a homogeneous cosmology, already
exists \cite{bib:gr-qc2006..07...111G} and it
would be desirable to apply it to the
non-singular baby-universe creation also.
Finally it would be also interesting to consider
the possibility of producing baby-universes
at the TeV scale in theories with large
compact extra-dimensions.

\section*{Acknowledgements}

We would like to thank H. Ishihara and
J. Portnoy for conversations.

\end{document}